\documentstyle[floats,aps,eqsecnum,graphics]{revtex}
\begin{document}
\twocolumn[\hsize\textwidth\columnwidth\hsize\csname@twocolumnfalse%
\endcsname
\draft

\title{Correlation function algebra for inhomogeneous fluids}
\author{A.\ O.\ Parry and P.\ S.\ Swain\cite{email}}
\address{Department of Mathematics, Imperial College\\ 180 Queen's Gate, London SW7 2BZ\\ United Kingdom.}
\maketitle

\begin{abstract}
We consider variational (density functional) models of fluids confined in parallel-plate geometries (with walls situated in the planes $z=0$ and $z=L$ respectively) and focus on the structure of the pair correlation function $G({\bf r}_1,{\bf r}_2)$. We show that for {\it local} variational models there exist two non-trivial identities relating both the transverse Fourier transform $G(z_\mu, z_\nu;{\bf q})$ and the zeroth moment $G_0(z_\mu,z_\nu)$ at different positions $z_1$, $z_2$ and $z_3$. These relations form an algebra which severely restricts the possible form of the function $G_0(z_\mu,z_\nu)$. For the common situations in which the equilibrium one-body (magnetization/number density) profile $m_0(z)$ exhibits an odd or even reflection symmetry in the $z=\frac{L}{2}$ plane the algebra simplifies considerably and is used to relate the correlation function to the finite-size excess free-energy $\gamma(L)$. We rederive non-trivial scaling expressions for the finite-size contribution to the free-energy at bulk criticality and for systems where large scale interfacial fluctuations are present. Extensions to non-planar geometries are also considered.
\end{abstract}

\vskip2pc]

\section{Introduction}
\label{sec:1}
In this paper we discuss the structure of correlation functions for fluids (or Ising-like magnets) adsorbed at walls and confined in parallel-plate (thin film), cylindrical and spherical geometries. Such systems have been extensively studied in recent years to assess the influence that surface and finite-size effects have on bulk phase coexistence and criticality. Here we are not primarily interested in the phenomenology but seek rather to understand whether a central assumption crucial to a broad class of widely used variational (density functional) models has any consequences for the behaviour of the correlation function $G({\bf r}_1,{\bf r}_2)$. Specifically, we shall show that for {\it local} functional models there exist very general relations which may be viewed as an algebra restricting the possible structure of $G({\bf r}_1,{\bf r}_2)$. These turn out to be particularly powerful when the one-body (magnetization/density) profile has an odd/even reflection symmetry (as is often the case) and enable us to re-derive non-trivial scaling laws for the finite-size contribution to the free-energy for systems close to the critical point \cite{fisher1,privman} or that have large scale interfacial fluctuations \cite{parry1,parry2}.

The starting point of our analysis is an appropriate variational model for the free-energy. For almost all of our article we will consider grand potential-like functionals $\Omega[m({\bf r})]$, where $m({\bf r})$ denotes the appropriate order parameter or local density variable. Thus for fluid systems $m({\bf r})$ corresponds to the number density although we shall adopt a magnetic notation and will often refer to $m({\bf r})$ as the local magnetization or spin density. The grand potential functional is written \cite{rowlinson,evans1,evans2}
\begin{equation}
\Omega[m({\bf r})] = {\cal F}[m({\bf r})] - \int m({\bf r}) h({\bf r}) d{\bf r}
\end{equation}
where ${\cal F}[m({\bf r})]$ is the Helmholtz free-energy functional and $h({\bf r})$ is the local magnetic field at point ${\bf r}$. In a fluid context $h({\bf r})$ may be identified with $\mu -V_{\rm ext}({\bf r})$ where $\mu$ is the chemical potential and $V_{\rm ext}({\bf r})$ is the external field generated by the confining walls. Minimization of $\Omega[m({\bf r})]$ with respect to magnetization configurations recovers the thermodynamic grand potential $\Omega$ \cite{rowlinson,evans1,evans2}.
\begin{eqnarray}
\Omega &=& \min \Omega[m({\bf r})] \nonumber \\
&=& \Omega[m_0({\bf r})]
\end{eqnarray}
where $m_0({\bf r})$ is the equilibrium magnetization profile. 

The connected correlation function $G({\bf r}_1,{\bf r}_2)$ is defined in the usual way by
\begin{equation}
G({\bf r}_1,{\bf r}_2) = \langle m({\bf r}_1) m({\bf r}_1) \rangle -\langle m({\bf r}_1) \rangle  \langle m({\bf r}_2) \rangle 
\end{equation}
and satisfies the Ornstein-Zernike integral equation \cite{rowlinson}
\begin{equation}
\int d{\bf r}' C({\bf r}_1,{\bf r}') G({\bf r}',{\bf r}_2) = \delta({\bf r}_1-{\bf r}_2) \label{OZ}
\end{equation}
where the direct correlation function $C({\bf r}_1,{\bf r}_2)$ is defined by 
\begin{equation}
C({\bf r}_1,{\bf r}_2)= \frac{1}{k_B T} \frac{\delta^2 F[m({\bf r})]}{\delta m({\bf r}_1) \delta m({\bf r}_2)} \label{C}
\end{equation}
and is evaluated at equilibrium.

One of the chief merits of the variational approach is the prescription for calculating $G({\bf r}_1,{\bf r}_2)$ via the direct correlation function route embodied in (\ref{OZ}) and (\ref{C}). Often the fluid has a symmetry which makes the task simpler. For example, in planar parallel-plate geometries the one-body profile is a function of one coordinate, $z$ say, and it is convenient to Fourier transform $G({\bf r}_1,{\bf r}_2)$ and $C({\bf r}_1,{\bf r}_2)$ with respect to the transverse displacement vector ${\bf y}_{12}$ of the two points ${\bf r}_1$ and ${\bf r}_2$. Thus we define
\begin{equation}
G(z_1,z_2;q) = \int d{\bf y}_{12} \exp(i {\bf q}.{\bf y}_{12}) G({\bf r}_1, {\bf r}_2) 
\end{equation}
and $C(z_1,z_2;q)$ similarly, which from (\ref{OZ}) satisfy 
\begin{equation}
\int dz_3 C(z_1,z_3;q) G(z_3,z_2;q) = \delta(z_1-z_2) \label{OZ2}
\end{equation}
We shall be particularly interested in the zeroth moment corresponding to $q=0$ and define (in standard notation)
\begin{equation}
G_0(z_1,z_2) \equiv G(z_1,z_2;0) \label{goo}
\end{equation}

For the planar parallel-plate geometry the finite-size contribution to the grand potential $\Omega$ is conveniently measured by the surface excess quantity
\begin{equation}
\gamma(L) \equiv \frac{\Omega - \omega_b V}{A}
\end{equation}
where $V\equiv AL$ is the volume of the thin film of width $L$ and area $A$. Here $\omega_b$ is the bulk grand potential density. The quantity $\gamma(L)$ may be regarded as the finite-size dependent surface tension (excess free-energy per unit area) of the system. Differentiating with respect to $L$ yields the solvation force \cite{ep}
\begin{equation}
f_s(L) \equiv -\frac{\partial \gamma(L)}{\partial L}
\end{equation}
which is a useful measure of phase behaviour in the confined fluid. The most important conclusions of this paper concern the derivation of elegant relations between the correlation function $G_0(z_1,z_2)$ and (derivatives of) the free-energy $\gamma(L)$.

Of course the exact density functional for non-trivial (interacting) three dimensional systems is not known and approximate models must be considered for the purposes of calculation. The simplest approach (and one which has given invaluable insights into a number of problems \cite{evans1}) is to assume that the Helmholtz functional is local so that we may write
\begin{eqnarray}
\Omega[m({\bf r})] & = & \int d{\bf y} \int_0^L dz \left\{ {\cal L}^{(b)}(m({\bf r}),\nabla m({\bf r})) \right. \nonumber \\
& & + \delta(z) \phi_1(m({\bf r})) +\delta(z-L) \phi_2(m({\bf r})) \Bigr \} \label{F}
\end{eqnarray}
where ${\cal L}^{(b)}(m,\nabla m)$ is an appropriate bulk free-energy density while $\phi_i(m)$ models (short ranged) interactions with the walls situated in the planes $z=0$ and $z=L$, say. Most often ${\cal L}^{(b)}(m,\nabla m)$ is further approximated by a Landau expansion \cite{evans1}
\begin{equation}
{\cal L}^{(b)}(m,\nabla m) = \frac{1}{2} (\nabla m)^2 + \frac{r}{2!} m^2 + \frac{u}{4!} m^4-hm \label{landau}
\end{equation}
where, for fluid systems $m$ must now be interpreted as the number density relative to the bulk critical value. The inadequacies of the Landau-type approach near the bulk critical temperature are well known though more general expressions for ${\cal L}^{(b)}(m,\nabla m)$ can account for non-classical critical exponents in a phenomenological manner \cite{eight}. Such, generalised, local free-energy functionals have provided a profitable way of deriving critical exponent and scaling relations, although recently, it has been argued that local entropy-like functionals may well be better candidates for describing critical effects \cite{mikheev}. As well as being restricted to systems with short ranged forces the local approximation is also inadequate for describing the pronounced oscillations in the density profile that occur when a high density fluid is confined in a narrow geometry \cite{evans1}. Nevertheless due to the continued wide spread use of local functionals we believe that it is worth considering in detail what restrictions the local approximation places on the structure of correlation functions.

To address this question we borrow recent results and methods developed for analysing correlation functions at wetting transitions which have led to the introduction of coupled effective Hamiltonians \cite{boulter,parry3,parry4}. While we shall not be particularly concerned with the wetting transition in this article, our analysis involves the further development and generalization of the stiffness-matrix formalism present in \cite{parry3} and \cite{parry4}. The essence of the method is to consider the properties of constrained functionals constructed by partial minimization of the grand potential functional $\Omega[m]$. The Ornstein-Zernike equation is then recast in a convenient matrix representation which allows us to exploit the separable properties of the constrained functionals. We emphasise here that the partial minimisation procedure is exact in the present variational formalism and contrasts with the analogous result in the effective Hamiltonian theory of wetting which arises due to a saddle point approximation \cite{fisher2}. The present analysis also differs from the coupled effective Hamiltonian theory since it is necessary to consider the properties of constrained functionals of three (or more) collective coordinates rather than just two.

For the greater part of our article, embodying Sec.\ \ref{sec:2}-\ref{sec:52}, we deal with planar inhomogeneous fluids described by local functionals of the form (\ref{F}) but leave ${\cal L}^{(b)}$ and $\phi_i$ arbitrary. Our presentation is as follows: in Sec.\ \ref{sec:21} we recall the basic strategy of the stiffness-matrix formalism considering constrained functionals of $N$ collective coordinates and introducing the $\frac{N(N-1)}{2}$ structure factor matrix elements $S_{\mu \nu}(q)$ (with $1 \leq \mu,\nu \leq N$). Then in Sec.\ \ref{sec:22} we show how the local character of $\Omega[m]$ necessarily leads to two sets of algebraic relations relating the $S_{\mu \nu}(q)$ and $S_{\mu \nu}(0)$ at any three planes $z_1, z_2$ and $z_3$. This means that for arbitrary positions $z_1 \leq z_2 \leq z_3$ the zeroth moments $G_0(z_\mu,z_\nu)$ satisfy two non-trivial identities. We illustrate this in Sec.\ \ref{sec:3} for the simplest possible scenario corresponding to fluid adsorbed at a single wall before turning our attention to the much more interesting problem of fluids confined in a parallel-plate geometry. For this case, if we consider examples in which the equilibrium one-body profile exhibits a reflection symmetry in the $z=\frac{L}{2}$ plane, the algebra simplifies considerably. Even symmetry corresponds to profiles satisfying $m_0(z)=m_0(L-z)$ and occurs if both confining walls are identical. For this case one of our relations readily re-derives an exact result due to Henderson \cite{henderson} valid for fluids (interacting with quite arbitrary interatomic forces) confined between perfectly hard walls. On the other hand profiles satisfying $m_0(z)=-m_0(L-z)$ correspond to odd symmetry and occur in Ising-like systems with competing surface fields. Such geometries have attracted considerable attention in recent years (see \cite{parry1,parry2,binder1,binder2} for example and references therein) due to presence of strong interfacial finite-size effects and novel symmetry breaking mechanisms. 

For both even and odd systems the correlation function relations severely restrict the form of the zeroth moment and we are able to derive a very elegant expression for $G_0(z_1,z_2)$ in terms of $G_0(z_1,z_1), G_0(z_2,z_2), m'_0(z)$ and $\gamma(L)$. Using this approach we rederive non-trivial scaling laws at the bulk critical point and when large scale interfacial fluctuations are present in the confining geometry. Sec.\ \ref{sec:5} concludes our discussion of the planar system where we consider the continuum ($N\rightarrow \infty$) limit of the relations. We then turn our attention to non-planar geometries for which the stiffness-matrix formalism  has not yet be fully developed. Nevertheless we show that for Landau-type models of inhomogeneous fluids in cylindrical and spherical symmetries it is possible to define analogous matrix elements $S_{\mu \nu}$ which satisfy the same algebraic relations as those in the planar geometry.

We conclude our article with a summary of our main results and make some remarks about other types of variational model.

\section{Correlation function relations}
\label{sec:2}
\subsection{Constrained functionals}
\label{sec:21}
Consider a fluid confined between two planar walls (of infinite transverse area $A$) situated in the planes $z=0$ and $z=L$. We consider the most general situation in which the walls are not identical and make no assumption about the preferential adsorption at each surface. Now consider an ordered arrangement of $N$ planes (labelled by $\alpha=1,\ldots,N$) corresponding to positions $z_1,z_2,\ldots,z_N$ and denote the equilibrium  magnetization 
\begin{equation}
m_0(z_\alpha)=m^X_\alpha \label{zalpha}
\end{equation} 
at each (see Fig.\ 1).
\begin{figure}[h]
\begin{center}
\scalebox{0.5}{\includegraphics{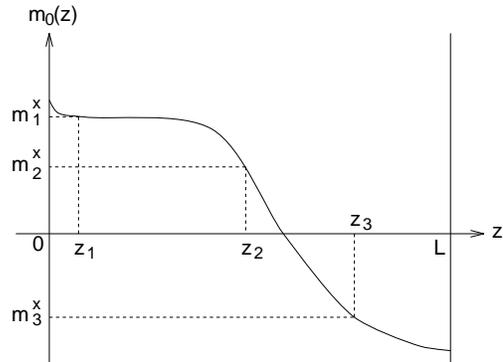}}
\caption{Schematic illustration of the equilibrium profile $m_0(z)$ for a thin film geometry. The equilibrium positions of surfaces of fixed magnetization $m^X_1,m^X_2$ and $m_3^X$ are shown.}
\end{center}
\end{figure}
We next introduce a constrained minimization of the functional $\Omega[m({\bf r})]$ which may be viewed as variational prescription for integrating out degrees of freedom except for those infinitesimally close to the planes. To this end suppose that $\Omega[m]$ is minimized subject to the condition that the surfaces described by the position variables $\ell_\alpha({\bf y})$ are contours of fixed magnetization $m_\alpha^X$
\begin{equation}
m(\ell_\alpha({\bf y}))=m^X_\alpha \mbox{\ \ \ \ \ $\forall \alpha$} \label{crossing}
\end{equation}
Thus we define a constrained functional \cite{fisher2}
\begin{equation}
H_N[ \{ \ell_\alpha({\bf y}) \} ; \{ z_\alpha \} ]= \overline{\min} \Omega[m({\bf r})] \label{inter}
\end{equation}
where the bar denotes the crossing criterion constraint (\ref{crossing}). By construction the equilibrium position of the surfaces of fixed magnetization $m^X_\alpha$ is $z_\alpha$ (see (\ref{zalpha})) and setting $\ell_\alpha({\bf y})=z_\alpha$ in (\ref{inter}) recovers the thermodynamic potential $\Omega$. To make connection with the correlation functions we need to consider the properties of the constrained functional in the vicinity of the global minimum. To this end we suppose that for small translations and fluctuations we may write \cite{parry3}
\begin{eqnarray}
H_N[\{\ell_\alpha({\bf y})\};\{z_\alpha \}] &=& \int d{\bf y} \left[ \frac{1}{2} \Sigma_{\mu \nu} (\{\ell_\alpha\};\{z_\alpha \}) \nabla \ell_\mu.\nabla \ell_\nu \right. \nonumber \\
& &  + W_N(\{\ell_\alpha\};\{z_\alpha \}) \Bigr ] \label{i2}
\end{eqnarray}
where the $\Sigma_{\mu \nu}$ constitute the elements of the stiffness-matrix ${\bf \Sigma}$ for this particular choice of $\{m^X_\alpha\}$ (or equivalently $\{ z_\alpha \}$, see (\ref{zalpha})). In writing (\ref{i2}) we have (for convenience) omitted to include terms related to the curvatures $\nabla^2\ell_\alpha$ etc.\ which may be accounted for using an appropriate rigidity matrix ${\bf K}$. Also in contrast to the analogous effective Hamiltonian expressions there is no need to specify the momentum cut-offs for each field. Obviously different choices of the $\{z_\alpha\}$ will result in different binding potentials $W_N$ and stiffness-matrices ${\bf \Sigma}$. For this reason it is sometimes convenient to speak of a continuous set of functionals $\{H_N(\{\ell_\alpha\};\{z_\alpha \}) \}$ whose elements are distinguished by the particular choice of $\{z_\alpha\}$ or equivalently $\{m^X_\alpha\}$. Connection with the correlation function $G(z_1,z_2;q)$ is made with an appropriate finite dimensional matrix representation of the Ornstein-Zernike integral equation (\ref{OZ}), \cite{parry4}. The analogue of the direct correlation function  is an $N \times N$ matrix with elements (setting $k_B T=1$)
\begin{equation}
\bar{C}_{\mu \nu} ({\bf y}_{12};\{ z_\alpha \}) =  \left. \frac{\delta^2 H_N(\{\ell_\alpha\};\{z_\alpha \})}{\delta \ell_\mu({\bf y}_1) \ell_\nu ({\bf y}_2)} \right|_{\ell_\mu=z_\mu} \label{Ca}
\end{equation}
which we Fourier transform to exploit the translational invariance. The wave vector expansion of the matrix ${\bf C}(q;\{z_\alpha \})$ is rather elegant
\begin{eqnarray}
{\bf C}(q;\{z_\alpha \}) & = & {\bf C}_0(\{z_\alpha \}) + q^2 {\bf \Sigma}( \{z_\alpha \};\{z_\alpha \}) \nonumber \\
& &  + q^4 {\bf K}( \{z_\alpha \};\{z_\alpha \}) + \ldots \label{exp}
\end{eqnarray}
and is central to the utility of the stiffness-matrix formalism. The zeroth term ${\bf C}_0$ is the matrix of curvatures
\begin{equation}
{\bf C}_0(\{z_\alpha \}) = \left[ \begin{array}{ccccc} 
\partial^2_{11} & \partial^2_{12} & . & . & \partial_{1N} \\
. & \partial^2_{22} & . & . & .  \\
\partial^2_{N1} & . & . & . & .  \end{array} \right] W_N( \{\ell_\alpha \};\{z_\alpha\}) \label{Cb}
\end{equation}
where $\partial^2_{\mu \nu} = \frac{\partial^2}{\partial \ell_\mu \partial \ell_\nu}$ and is evaluated at equilibrium $\ell_\alpha =z_\alpha$. The correlation function $G(z_\mu,z_\nu;q)$ then follows as \cite{parry3,parry4}
\begin{equation}
G(z_\mu,z_\nu;q) = m'_0(z_\mu) m'_0(z_\nu) S_{\mu \nu}(q; \{z_\alpha\})
\end{equation}
where the $S_{\mu \nu}$ are the elements of the (symmetric) structure factor matrix satisfying
\begin{equation}
{\bf S} (q;\{ z_\alpha \}) {\bf C} (q;\{ z_\alpha \}) = {\bf I} \label{Smatrix}
\end{equation}
with ${\bf I}$ the identity matrix. These equations have previously been used to study mean-field correlation functions at wetting transitions \cite{parry4}. In that context the $H_N( \{\ell_\alpha\};\{z_\alpha\})$, correspond to coupled effective Hamiltonians and the minimization condition arises as a saddle point approximation to a functional integral. Here we emphasise that the formulation is an {\it exact} prescription for calculating correlation functions provided one begins with the appropriate variational functional.

\subsection{Local functionals}
\label{sec:22}
To begin consider the zeroth moment $G_0(z_\mu,z_\nu)$ for which we only need the binding potential. In order to derive the correlation function relations it is sufficient to consider the properties of `three field' functionals $H_3[\ell_1,\ell_2,\ell_3]$ . as working with $N>3$ does not give any new results. The essential observation leading to the correlation function relations is that due to the local character of the grand potential functional (\ref{F}) the binding potentials necessarily have a separable form (see Appendix A)
\begin{eqnarray}
& W_3(\{\ell_\alpha({\bf y})\};\{z_\alpha \}) =  V_1(\ell_1; \{z_\alpha \}) & \nonumber \\
& + V_2(\ell_2-\ell_1; \{z_\alpha \}) + V_3(\ell_3-\ell_2; \{z_\alpha \}) & \nonumber \\
& + V_4(\ell_3; \{z_\alpha \}) &
\end{eqnarray}
where the functions $V_i$ depend on the particular choice of variational model and may be considered unknowns. From (\ref{Cb}) it is a trivial exercise to calculate the matrix ${\bf C}_0(\{z_\alpha\})$  and note the following general properties
\begin{eqnarray}
C_{13}(0;\{z_\alpha\}) &=&0 \label{con1} \\
C_{12}(0;\{z_\alpha\}) + C_{22}(0;\{z_\alpha\}) + C_{23}(0;\{z_\alpha\}) &=& 0 
\end{eqnarray}
In turn these impose conditions, via (\ref{Smatrix}), on the structure factor matrices ${\bf S}(0;\{z_\alpha\})$ (hereafter we drop the explicit $\{ z_\alpha \}$ dependence).
\begin{equation}
\left| \begin{array}{ll} 
S_{12}(0) & S_{13}(0) \\
S_{22}(0) & S_{23}(0) \end{array} \right| =0 \label{ratio}
\end{equation}
and 
\begin{equation}
\left| \begin{array}{ll} 
S_{12}(0) & S_{23}(0) \\
S_{13}(0) & S_{33}(0) \end{array} \right| + 
\left| \begin{array}{ll} 
S_{11}(0) & S_{12}(0) \\
S_{13}(0) & S_{23}(0) \end{array} \right| =
\left| \begin{array}{ll} 
S_{11}(0) & S_{13}(0) \\
S_{13}(0) & S_{33}(0) \end{array} \right| \label{hello}
\end{equation}

For $q\neq0$ it is necessary to consider the properties of the stiffness-matrix and rigidity etc.\ appearing in the expansion (\ref{exp}). Nevertheless because of the local character of $\Omega[m]$, variation of $\ell_1({\bf y})$, say, does not effect the constrained magnetization in the region $z \ge \ell_2({\bf y})$ (see Appendix A). Consequently the $\Sigma_{13}$ element of the stiffness-matrix vanishes (as does $K_{13}$ the corresponding element in the rigidity matrix) leading to
\begin{equation}
C_{13}(q;\{ z_\alpha \}) =0 \mbox{\ \ \ \ \ $\forall q$}
\end{equation}
which is clearly the generalization of (\ref{con1}) and leads to the first of our identities
\begin{equation}
S_{12}(q) S_{23}(q) = S_{22}(q) S_{13}(q) \label{sPRODUCT}
\end{equation}
consistent with (\ref{ratio}) when $q=0$. The second identity is restricted to $q=0$ and is given in (\ref{hello}) above. This may be profitably rewritten as
\begin{eqnarray}
& \left[ S_{11}(0)-S_{12}(0) \right] \left[ S_{33}(0)-S_{23}(0) \right] = & \nonumber \\
& \left[ S_{13}(0)-S_{12}(0) \right]  \left[ S_{13}(0) - S_{23}(0) \right] & \label{other}
\end{eqnarray}
If we were to consider an $N$-field constrained functional $H_N[ \{\ell_\alpha \}; \{z_\alpha \}]$ the $S_{\mu \nu}(q)$ are related by the same relations applied to any ordered triplet of planes located at $z_\alpha \le z_\beta \leq z_\gamma$, i.\ e.\
\begin{equation}
S_{\alpha \beta}(q) S_{\beta \gamma}(q) = S_{\beta \beta}(q) S_{\alpha \gamma}(q) \label{PRODUCT}
\end{equation}
and 
\begin{eqnarray}
& \left[ S_{\alpha \alpha}(0)-S_{\alpha \beta}(0) \right] \left[ S_{\gamma \gamma}(0)-S_{\beta \gamma}(0) \right] = & \nonumber \\
& \left[ S_{\alpha \gamma}(0)-S_{\alpha \beta}(0) \right]  \left[ S_{\alpha \gamma}(0) - S_{\beta \gamma}(0) \right] &  \label{SUM}
\end{eqnarray}
Clearly (\ref{PRODUCT}) and (\ref{SUM}) define an algebra satisfied by the $\frac{N(N-1)}{2}$ quantities $S_{\mu \nu}(q)$ (for each $q$). In terms of the pair correlation function $G(z_\mu,z_\nu;q)$ they read
\begin{equation}
G(z_1,z_2;q) G(z_2,z_3;q) = G(z_2,z_2;q) G(z_1,z_3;q) \label{ratio1}
\end{equation}
and
\begin{eqnarray}
& \frac{ m'_0(z_2) G_0(z_1,z_1)-m'_0(z_1) G_0(z_1,z_2)}{ m'_0(z_2) G_0(z_1,z_3)-m'_0(z_3) G_0(z_1,z_2)} = & \nonumber \\
&  \frac{ m'_0(z_2) G_0(z_1,z_3)-m'_0(z_1) G_0(z_2,z_3)}{ m'_0(z_2) G_0(z_3,z_3)-m'_0(z_3) G_0(z_2,z_3)} & \label{otherG}
\end{eqnarray}
for all $0 \le z_1 \le z_2 \le z_3 \le L$. We emphasise that these relations are valid for all local variational models of the form (\ref{F}). As we shall see, taken together they restrict the form of the zeroth moment $G_0(z_1,z_2)$ and allow us to derive elegant expressions relating $G_0$ to the force of solvation $f_s(L)$. Before we do this we consider the case of fluid adsorption at a single wall for which the algebra (\ref{PRODUCT}) and (\ref{SUM}) has a trivial solution.

\section{A simple case: fluid adsorption at a single wall}
\label{sec:3}
We wish to show how for a semi-infinite system the algebra conditions, (\ref{PRODUCT}) and (\ref{SUM}), with $q=0$ are met. To this end we only need to consider the properties of the binding potential $W_N(\{\ell_\mu \})$. For later purposes it is convenient to position the wall in the plane $z=z_0$ (fixed). Clearly the equilibrium profile $m_0(z;z_0)$ is a function of $z-z_0$ only so that the partial derivatives satisfy $\partial_z m_0(z;z_0) = -\partial_{z_0} m(z;z_0)$. Due to the local character of the grand potential function the $N$-field binding potential may be written
\begin{eqnarray}
W_N(\{\ell_\mu \}) & = & V_1(\ell_1)+V_2(\ell_2-\ell_1)+V_3(\ell_3-\ell_2) \nonumber \\
& & + \ldots + V_N(\ell_N-\ell_{N-1})
\end{eqnarray}
The simplifying feature here (compared to the parallel-plate geometry) is that the final collective coordinate $\ell_N({\bf y})$ only enters through one (unspecified) partial binding potential function $V_N(x)$. The curvature matrix has the tridiagonal form

\begin{equation}
{\bf C}_0= \left( \begin{array}{cccccc}
	V_1''+V_2'' & -V_2''   & 0 & . &.  & . \\
	-V_2'' & V_2''+V_3''  & . & . & . \\
	0 & -V_3''  & . & . & . & . \\
	0 & 0  & . & . & . &.  \\
	0 & . & . & . & -V''_{N-1} & 0 \\
	. & .  & . & . & V''_{N-1} + V''_{N} & -V_N''\\
	. & . & .  & 0  & -V_N'' & V''_N
\end{array} \right)
\end{equation}

where we have written $V''_\alpha \equiv \frac{\partial^2 V_\alpha(x)}{\partial x^2}$ (evaluated at equilibrium $\ell_\mu=z_\mu$). The inverse matrix ${\bf S}_0$ has a remarkably simple block structure

\begin{equation}
{\bf S}_0= \left( \begin{array}{ccccccc}
	S_{11} & S_{11} & S_{11} & S_{11} & .  & . & S_{11} \\
	S_{11} & S_{22} & S_{22} & S_{22} & .  & . & . \\
	S_{11} & S_{22} & S_{33} & S_{33} & .  & . & . \\
	S_{11} & S_{22} & S_{33} & .  & . & .  & . \\
	. & S_{22} & S_{33} & . & . & .  & S_{N-2,N-2} \\
	. & . & . & . &  . & S_{N-1,N-1} & S_{N-1,N-1}\\
	S_{11} & . & . & . & . & S_{N-1,N-1} & S_{NN} 
\end{array} \right)
\end{equation}

with
\begin{equation}
S_{\mu \mu} \equiv S_{\mu \mu}(0) = \sum_{\alpha=1}^\mu\frac{1}{V_\alpha''}
\end{equation}
The point to notice here is that
\begin{equation}
S_{\mu \nu}(0) = S_{\mu \mu}(0) \mbox{\ \ \ \ \ $\forall \nu \ge \mu$} \label{semi}
\end{equation}
which is indeed a solution of the algebraic conditions (\ref{PRODUCT},\ref{SUM}). Thus we are led to the conclusion that within all local variational models of fluid adsorption at a single wall the pair correlation function is of the form
\begin{equation}
G_0(z_1,z_2) = \partial_z m_0(z_1;z_0) \partial_z m_0(z_2;z_0) F \left( \min[z_1,z_2] \right)
\end{equation}
or equivalently
\begin{equation}
G_0(z_1,z_2) = \partial_{z_0} m_0(z_1;z_0) \partial_{z_0} m_0(z_2;z_0) F \left( \min[z_1,z_2] \right) \label{general}
\end{equation}
where $F(x)$ is an unknown function.

This prediction is supported by the explicit Landau theory result for a semi-infinite system (i.\ e.\ if we assume that ${\cal L}^{(b)}$ is given by (\ref{landau})) which is known to be \cite{parry3,parry4,parry5,evans3} 
\begin{equation}
G_0(z_1,z_2) = m'_0(z_1)m'_0(z_2)\left( \alpha_0 + \int_{z_0}^{\min (z_1,z_2)} \frac{dz}{m'_0(z)^2} \right) \label{LPl}
\end{equation}
with
\begin{equation}
\alpha_0^{-1} = m'_0(0)[\phi''_1 m'_0(0) - m''_0(0)] 
\end{equation}
where we have abbreviated $m'_0(z) = \partial_z m_0(z;z_0)$ and $\phi_1''(m)=\frac{d^2 \phi_1}{dm^2}$ .

Equation (\ref{semi}) is also consistent with an exact result due to Henderson and van Swol \cite{hvs} who have derived an exact statistical mechanical sum rule for fluid adsorption at a pure hard wall i.\ e.\ an external potential
\begin{equation}
V_{HW}(z) = \left\{ \begin{array}{l} \infty \mbox{\ \ \ \ \ for $z<0$} \\ 0 \mbox{\ \ \ \ \ \ for $z>0$} \end{array} \right. \label{hw}
\end{equation}
which simply confines the fluid to the half plane $z \ge0$. 

For this case it is possible to show that
\begin{equation}
G_0(0,z)=\rho'_0(z) \mbox{\ \ \ \ \ for hard walls}
\end{equation}
with $\rho_0(z)$ ($\Leftrightarrow m_0(z)$) the equilibrium local number density. This is in agreement with (\ref{semi}) for the special case $z_\mu=0$ equivalent to
\begin{equation}
G_0(0,z) \propto m_0'(z)
\end{equation}
with an unknown constant of proportionality which presumably depends on the choice of surface interaction term $\phi_1(m)$.

While the semi-infinite solution (\ref{semi}) is not particularly interesting it does represent the correct $L\rightarrow \infty$ limit of the algebra pertinent to fluids confined in parallel-plate geometries. The analysis for this case is much richer and is discussed in the next section.

\section{Parallel plate geometries}
\label{sec:4}
\subsection{Preliminary remarks}
\label{sec:41}
There are two cases when the correlation function relations (\ref{PRODUCT}) and (\ref{SUM})---with $q=0$, simplify further and yield elegant expressions for $G_0(z_1,z_2)$. These occur in parallel-plate geometries in which the one-body profile $m_0(z)$ exhibits a simple reflection symmetry. For example if both walls are identical it necessarily follows that $m_0(z)$ is an {\it even} function about the plane of symmetry $z=\frac{L}{2}$. These systems have traditionally attracted the most attention in the literature where the shift of the bulk critical point and first-order phase boundary (capillary condensation) are of interest. The finite-size phase diagram and representative profiles for a typical parallel-plate geometry are shown schematically in Fig.\ 2.
\begin{figure}[h]
\begin{center}
\scalebox{0.5}{\includegraphics{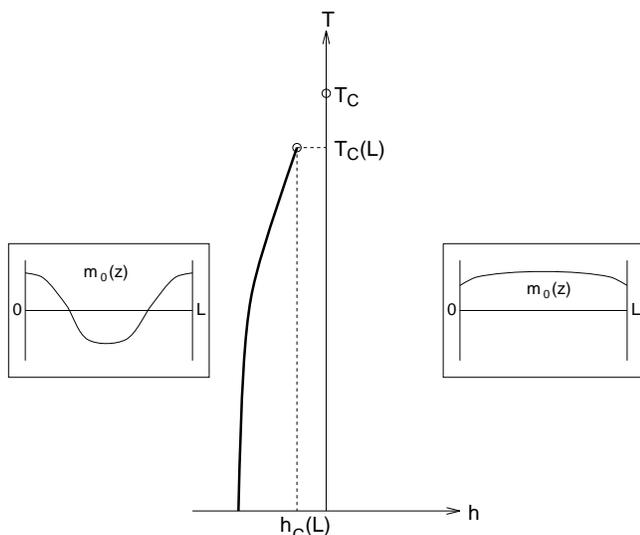}}
\caption{Schematic phase diagram of a thin film Ising-like magnet with positive surface fields $h_1=h_2$. Representative magnetization profiles, shown inset have an even reflection symmetry. The locus of first-order phase transitions (often referred to as capillary condensation) is shifted away from the zero bulk field ($h=0$) line. The shifted capillary critical point occurs at $T_C(L)$ and $h_C(L)$.}
\end{center}
\end{figure}
However more recently, examples in which the profile exhibits an {\it odd} reflection symmetry have also drawn considerable interest. Such parities arise in Ising-like systems confined by walls which exert surface fields $h_1$ and $h_2$ of equal magnitude but opposite sign on the spins in the $z=0$ and $z=L$ planes respectively. The nature of the phase coexistence and criticality in the system is entirely different to that occurring for the case of identical walls. In particular the critical temperature $T_C(L)$ for finite $L$ is determined by length scales associated with wetting \cite{parry1} and is restricted to lie close to the wetting temperature $T_W$ of the semi-infinite system (see Fig.\ 3).
\begin{figure}[h]
\begin{center}
\scalebox{0.5}{\includegraphics{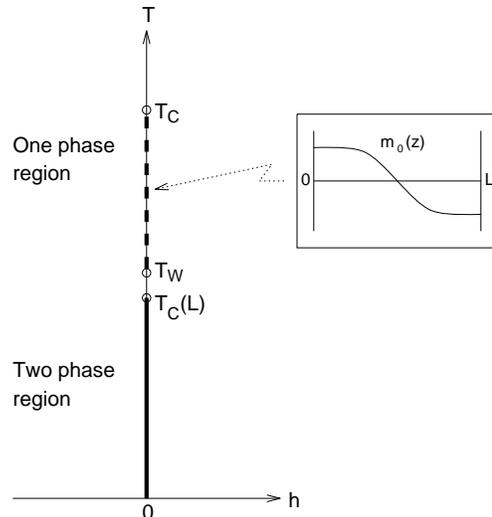}}
\caption{Schematic phase diagram of thin film Ising model with opposite surface fields $h_1=-h_2$. For temperatures $T \ge T_C(L)$ (and zero bulk field) the magnetization profiles have an odd reflection symmetry. In the temperature window $T_C \ge T \ge T_C(L)$ the profile resembles an interface located near the middle of the thin film and characterized by an extremely large interfacial correlation length $\xi_\parallel$. The location of the capillary critical point is determined by lengthscales pertinent to wetting and so $T_C(L)$ is less than but close to the critical wetting temperature $T_W$.}
\end{center}
\end{figure}
For temperatures $T>T_C(L)$ and $h=0$ the magnetization profile has the odd symmetry specified above. Of particular interest is the temperature window $T_C>T>T_W$ corresponding to the regime where the finite-size effects have suppressed bulk phase coexistence. The profile $m_0(z)$ for these temperatures resembles an up spin-down spin interface situated in the centre of the system. This interface is very weakly pinned by the confining walls and wanders almost freely in the finite-size geometry. The fluctuations associated with this wandering are extremely large leading to universal scaling behaviour for sufficiently low dimension \cite{parry2}. We shall return to this in Sec.\ \ref{sec:43} where we use the relations derived in Sec.\ \ref{sec:42} to calculate the singular contribution to the finite-size free-energy $\gamma(L)$ (for both even and odd systems).

To begin however, we make some prelimary remarks which will be useful for later purposes and allow us to make contact with other work. To establish our notation we note that the magnetization profile satisfies 
\begin{equation}
m'_0(z) = \pm m_0'(L-z)	
\end{equation}
for even ($-$) and odd ($+$) systems. Moreover we can also impose the additional symmetry requirement
\begin{equation}
G_0(z_1,z_1) = G_0(L-z_1,L-z_1)
\end{equation}
or equivalently
\begin{equation}
S_{11}(0)=S_{33}(0) \mbox{\ \ \ \ \ for $z_3=L-z_1$}
\end{equation}
Substitution into (\ref{SUM}) yields the `sum' rule
\begin{equation}
S_{12}(0)+S_{23}(0)=S_{11}(0)+S_{13}(0) \mbox{\ \ \ \ \  $z_3=L-z_1$} \label{sum}
\end{equation}
which compares to the `product' rule (\ref{PRODUCT})
\begin{equation}
S_{12}(0) S_{23}(0)=S_{22}(0) S_{13}(0)
\end{equation}
We can therefore anticipate that $S_{12}(0)$ and $S_{23}(0)$ must correspond to the roots of a single quadratic equation (see Sec.\ \ref{sec:42} below). In terms of the zeroth moment itself (\ref{sum}) implies
\begin{equation}
\frac{ G_0(z_1,z_2) \pm G_0(z_2,L-z_1)}{ G_0(z_1,z_1) \pm G_0(z_1,L-z_1)} = \frac{m'_0(z_2)}{m'_0(z_1)} \
\end{equation}
for even ($-$) and odd ($+$) systems and $z_1 \le z_2 \le L-z_1$. Further insight follows if we set $z_1=0$ to find
\begin{equation}
G_0(0,z)\pm G_0(z,L)=c m'_0(z) \label{hendy}
\end{equation}
where $c$ is an unknown constant (which may depend on $T,L,h,\ldots$). Note that the equilibrium magnetization profile is also a function of these variables.

It should be emphasized at this stage that this elegant relation follows as a necessary consequence of the local nature of the underlying variational model (\ref{F}). It is therefore encouraging to note that again it is consistent with an {\it exact} result due to Henderson \cite{henderson} who has considered statistical mechanical sum rules for a fluid with {\it arbitrary} intermolecular forces confined between two identical hard walls (i.\ e.\ {\it even} symmetry). Henderson supposed that the external potential could be written as a sum of two semi-infinite wall contributions
\begin{equation}
V_{\rm ext}(z) = V_\infty(z)+V_\infty(L-z)
\end{equation}
and derived a number of exact relations between integrals over $G_0(z_1,z_2)$ and one-body/thermodynamic quantities. For the specific case of hard walls $V_\infty=V_{HW}$ these simplify and in particular yield
\begin{equation}
G_0(0,z)-G_0(z,L)=\rho'_0(z) \mbox{\ \ \ \ \ for hard walls} \label{hendy1}
\end{equation}
with $\rho_0(z)$ ($\Leftrightarrow m_0(z)$) the equilibrium (local) number density. Clearly this is in agreement with our prediction (\ref{hendy}) and identifies the unknown constant $c$ as the universal value $c_{HW}=1$, for hard walls (\ref{hw}). Henderson \cite{henderson} also derives a relation between the pair correlation function and the free-energy for this system (recall that we have set $k_bT=1$)
\begin{equation}
-\frac{d^2 \gamma}{dL^2} = G_0(0,L) \mbox{\ \ \ \ \ for hard walls} \label{hendy2}
\end{equation}
In the next section we show how this (and more) can be derived from the correlation function algebra for more general wall potentials.

\subsection{Connection with the free-energy}
\label{sec:42}

For $q=0$ there are two identities satisfied by the $S_{\mu \nu}(0)$ at any three planes $z_1 \le z_2 \le z_3$ valid for arbitrary (short ranged) wall interactions $\phi_1$ and $\phi_2$. To develop the theory further beyond the elementary remarks made above, it is necessary to inquire what restrictions these conditions impose on the structure of $S_{\mu \nu}(0)$. Consider for example (\ref{PRODUCT}), with $q=0$. In order that this is satisfied for arbitrary choices of $z_1,z_2$ and $z_3$ it follows that the structure factor must have the form of an {\it ordered} product
\begin{equation}
S_{\mu \nu}(0)=y^-(z_\mu) y^+(z_\nu) \mbox{\ \ \ \ \ for $z_\mu \le z_\nu$} \label{breakdown}
\end{equation}
regardless of whether the profile $m_0(z)$ exhibits a reflection symmetry or not. Here $y^-(z)$ and $y^+(z)$ are unknown functions, the properties of which we need to determine. Note that the ordering condition $z_\mu \le z_\nu$ is of some importance here and is less restrictive than the assertion that $S_{\mu \nu}(0)$ is a separable function of $z_\mu$ and $z_\nu$. While it is clear that similar remarks also apply for $q\ne0$, we do not have a second relation which yields valuable information about $y^-(z)$ and $y^+(z)$. To proceed we substitute (\ref{breakdown}) into (\ref{SUM}) to find
\begin{eqnarray}
& y^+(z_1) y^-(z_3)-y^+(z_1) y^-(z_2)-y^+(z_2) y^-(z_3)= & \nonumber \\
& y^+(z_3) y^-(z_1)-y^-(z_2) y^+(z_3)-y^+(z_2) y^-(z_1) &
\end{eqnarray}
which can be differentiated with respect to $z_2$ yielding (for all $z_1$ and $z_3$)
\begin{eqnarray}
y^+(z_2)' & = & \left( \frac{y^+(z_3)-y^+(z_1)}{y^-(z_3)-y^-(z_1)} \right) y^-(z_2)'  \nonumber \\
\Rightarrow y^-(z_2)' & = & \beta y^+(z_2)' 
\end{eqnarray}
for constant $\beta$. This relation is valid for all $z_2$ and can be integrated to show that $y^-(z)$ and $y^+(z)$ are linearly related
\begin{equation}
y^+(z)= \alpha + \beta y^-(z) \label{kernellink}
\end{equation}
with $\alpha$ the constant of integration. By using (\ref{breakdown}) and (\ref{kernellink}), we can solve for $y^-(z_\mu)$, say, in terms of $S_{\mu \mu}(0)$ and then use (\ref{breakdown}) again to yield an expression for $S_{\mu \nu}(0)$. We find
\begin{equation}
K S_{\mu\nu}(0) =\left( 1\pm \sqrt{1-K S_{\mu \mu}(0)} \right) \left( 1\pm \sqrt{1-K S_{\nu \nu}(0)} \right) \label{require}
\end{equation}
where $K=-\frac{4 \beta}{\alpha^2}$ is a single undetermined constant. The $\pm$ signs do {\it not} correspond to the odd/even reflection symmetries mentioned in Sec.\ \ref{sec:41} since we have not yet specialised to these systems. Instead they refer to the positions of $z_\mu$ and $z_\nu$ relative to the maximum of $G_0(z,z)$, as we shall see below.

Before we turn to odd/even systems, for which we can explicitly calculate the constant $K$, we note that the general solution (\ref{require}) is consistent with remarks made earlier concerning the semi-infinite limit. The algebra for this case corresponds to the $K \rightarrow 0$ limit of (\ref{require}), with the appropriate choice of signs,
\begin{eqnarray}
\stackrel{\rm lim}{\scriptscriptstyle{K \rightarrow 0}} S_{\mu \nu}(0) &=& \stackrel{\rm lim}{\scriptscriptstyle{K \rightarrow 0}} \frac{1}{K} \left(1-\sqrt{1-K S_{\mu \mu}(0)} \right) \nonumber \\
& & \times \left(1+\sqrt{1+K S_{\nu \nu}(0)} \right) \nonumber \\
& = & S_{\mu \mu}(0) 
\end{eqnarray}
as indicated in (\ref{semi}).

Hereafter we specialise to odd ($+$) and even ($-$) systems. Rearranging (\ref{require}) and using the results of Sec.\ \ref{sec:41} it is straight forward to derive 
\begin{equation}
K_{\pm}=\frac{\pm 4G_0(0,L) m'_0(0)^2}{(G_0(0,0)\pm G_0(0,L))^2} \label{K}
\end{equation}
However it is also possible to relate $K_{\pm}$ to the excess free-energy $\gamma(L)$. We simply quote the result and refer the interested reader to appendix B for the details
\begin{equation}
K_\pm = 4 \frac{d^2 \gamma(L)}{dL^2} \label{k+-}
\end{equation}
For odd symmetric systems it is also straight forward to derive
\begin{equation}
K_+ = \frac{m_0'(\frac{L}{2})^2}{G_0(\frac{L}{2},\frac{L}{2})} \label{k+}
\end{equation}
which is particularly useful. Note that the right hand side of this relation can be identified as the reciprocal of the maximum value of $S_{\mu \mu}(0)$ as a function of $z_\mu$. In fact, using (\ref{breakdown}) and (\ref{kernellink}) to write $S_{\mu \mu}(0)$, and assuming analyticity at all points (true for odd systems only), simple differentiation with respect to $z_\mu$ shows the maximum value to be $S_{\mu \mu}(0)=-\frac{\alpha^2}{4 \beta}$, which is just $\frac{1}{K}$.

Thus we have established the following relationship between the correlation function and excess free-energy
\begin{equation}
\frac{d^2 \gamma}{dL^2} = \pm \frac{G_0(0,L) m'_0(0)^2}{[G_0(0,0) \pm G_0(0,L)]^2} \label{ohyes}
\end{equation}
which is once again (remembering (\ref{hendy})) consistent with Henderson's exact results (\ref{hendy1},\ref{hendy2}) for fluids confined between hard walls (\ref{hw}).

In this way we are led to a very elegant universal equation relating suitably scaled moments of the correlation function. Specifically, define the {\it dimensionless} quantities
\begin{equation}
\sigma_{\mu \nu} \equiv  \frac{4}{k_BT} \frac{d^2\gamma}{dL^2} \frac{G_0(z_\mu,z_\nu)}{m_0'(z_\mu)m_0'(z_\nu)} \label{dfy}
\end{equation}
where for completeness we have reinstated the Boltzmann factor. For both odd and even systems the value of these variables at any two planes $z_\mu$ and $z_\nu (\ge z_\mu)$ are related by 
\begin{equation}
\sigma_{\mu \nu} = \left( 1 \pm \sqrt{1-\sigma_{\mu\mu}} \right) \left( 1\pm \sqrt{1-\sigma_{\nu\nu}} \right) \label{sig}
\end{equation}
where the $\pm$ signs must be chosen appropriately (see Fig.\ 4). 
\begin{figure}[h]
\begin{center}
\scalebox{0.5}{\includegraphics{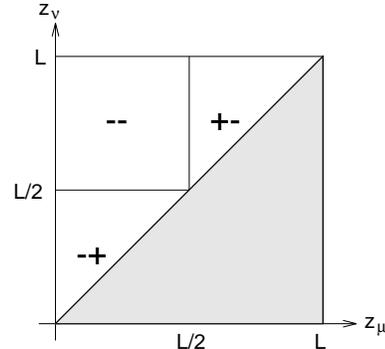}}
\caption{Choice of signs appearing in the first and second parenthesis, respectively in equation (4.24).}
\end{center}
\end{figure}
It is a straight forward exercise to check that this result for $\sigma_{\mu \nu}$ is consistent with the algebraic relations
\begin{eqnarray}
\sigma_{12} \sigma_{23} & = & \sigma_{22} \sigma_{13} \\
(\sigma_{11}-\sigma_{12}) (\sigma_{33}-\sigma_{23}) & = & (\sigma_{13}-\sigma_{12})(\sigma_{13}-\sigma_{23})
\end{eqnarray}
for $z_1 \le z_2 \le z_3$.

Making note of the sign of $K_\pm$ we see that even and odd systems are distinguished by the ranges of values of the $\sigma_{\mu \nu}$ variables. In particular
\begin{equation}
0 \le \sigma_{\mu \nu} \le 1 \mbox{ \ \ \ \ \ for odd systems}
\end{equation}
with $\sigma_{\mu \mu}=1$ for $z_\mu=\frac{L}{2}$. On the other hand in even systems where the force of solvation $f_s(L)$ between the plates is attractive we have
\begin{equation}
\sigma_{\mu \mu} \le 0 \mbox{ \ \ \ \ \ for even systems}
\end{equation}
and is not bounded from below. Note that $\sigma_{\mu \mu} \rightarrow -\infty$ as $z_\mu \rightarrow \frac{L}{2}$ due to the turning point in the even equilibrium magnetization profile (see Fig.\ 2). For both even and odd systems the variables $\sigma_{\mu \nu}$ and $\sigma_{\mu \mu}$ approach zero when $L\rightarrow \infty$ (for fixed temperature $T$) consistent with the semi-infinite limit $K_\pm \rightarrow 0$ mentioned earlier.

Equations (\ref{K}-\ref{sig}) are the main results of this paper and are obeyed by all {\it local} variational models. In the next section we shall demonstrate how they may be used to derive non-trivial scaling expressions for the fluctuation contribution to the free-energy $\gamma(L)$.

\subsection{Scaling of the finite-size free-energy}
\label{sec:43}
There are three different scenarios involving fluids confined in parallel-plate geometries for which the free-energy $\gamma(L)$ may be expected to exhibit singular, non-analytic behaviour arising from fluctuation related effects. These are considered separately below

\subsubsection{Odd and even systems at bulk criticality}
\label{sec:431}
Exactly at the bulk critical point $T=T_C$ and $h=0$ the correlation length in the finite-size system is only limited by the width $L$. This implies long-ranged (i.\ e.\ algebraic) behaviour in the finite-size free-energy $\gamma(L)$ which may be likened to the Casimir effect in quantum field theory \cite{barber}. The required non-classical (i.\ e.\ non-mean-field) scaling of $\gamma(L)$ may be derived from our equations in a number of ways. Perhaps the most elegant is to assert that due to the universal character (i.\ e.\ temperature and bulk field {\it independence}) of the scaled correlation function relation (\ref{sig}) the variables $\sigma_{\mu \nu}$ must exhibit finite-size scaling of the form
\begin{equation}
\sigma_{\mu \nu} \approx \Xi^{\pm} \left( \frac{z_\mu}{L}, \frac{z_\nu}{L} \right) \mbox{\ \ \ \ \ at $T=T_C$} \label{dfy2}
\end{equation}
where the scaling functions $\Xi^{\pm}(x,y)$ are universal \cite{barber} for all odd ($+$) and even ($-$) systems (restricting our attention to surface fields $h_1=\pm h_2(>0)$ and positive surface enhancement). This hypothesis may be compared with the definitions (\ref{dfy}) involving the free-energy. Standard finite-size arguments dictate that sufficiently far away from the wall the correlation function and magnetization profile behave (exactly at criticality) as \cite{fisher1,privman,barber}
\begin{equation}
G_0(z_1,z_2) \approx L^{\frac{\gamma}{\nu}-1} g^\pm \left( \frac{z_1}{L}, \frac{z_2}{L} \right) \label{gscale}
\end{equation}
and
\begin{equation}
m_0(z) \approx L^{-\frac{\beta}{\nu}} \Lambda^\pm \left( \frac{z}{L} \right) \label{mscale}
\end{equation}
where $g$ and $\Lambda$ are appropriate scaling functions, and $\gamma,\nu$ and $\beta$ are standard bulk critical exponents. Combining these with (\ref{dfy}), (\ref{dfy2}) yields
\begin{equation}
\gamma(L) \sim L^{1-\frac{(2-\alpha)}{\nu}} \mbox{\ \ \ \ \ at $T=T_C$} \label{scale}
\end{equation}
where we have used the Rushbrooke relation $2-\alpha=2\beta+\gamma$. Thus for $d>4$ we expect $\gamma(L) \sim L^{-3}$ while for $d<4$ we find $\gamma(L) \sim L^{-(d-1)}$ on invoking hyperscaling. This is precisely the predicted scaling behaviour of the surface excess free-energy \cite{fisher1,privman,barber} and serves to illustrate the generality of the relations derived in Sec.\ \ref{sec:42}. While they may be of restricted validity, applicable only to {\it local} variational model, they are not necessarily mean-field-like in character.

One may also derive (\ref{scale}) using (\ref{ohyes}) which implies
\begin{equation}
\frac{d^2 \gamma}{dL^2} \propto G_0(0,L) \mbox{\ \ \ \ \ as $L \rightarrow \infty$} \label{cool}
\end{equation}
at all temperatures. Using the `product' relation one has
\begin{equation}
G_0(0,L) = \frac{ G_0(0,\frac{L}{2})^2}{G_0 \left( \frac{L}{2},\frac{L}{2} \right)}
\end{equation}
and it is then straight forward to calculate the $L$ dependence of the right hand side of (\ref{cool}) at bulk criticality. The denominator follows from (\ref{gscale}) yielding $G_0 \left( \frac{L}{2},\frac{L}{2} \right) \sim L^{\frac{\gamma}{\nu}-1}$. In principle the scaling of $G_0(0,\frac{L}{2})$ could be calculated from the short distance expansion (valid for $\frac{z_1}{L} \ll 1$) of (\ref{gscale}). However it is easier to use the semi-infinite result $G_0(0,z) \propto m'_0(z)$ quoted in Sec.\ \ref{sec:3}. Combining these with the finite-size scaling of the profile (\ref{mscale}) it is natural to suppose
\begin{equation}
G_0(0,\frac{L}{2}) \sim L^{-\frac{\beta}{\nu}+1}  \mbox{\ \ \ \ \ at $T=T_C$}
\end{equation}
which recovers $\gamma(L) \sim L^{-\frac{2-\alpha}{\nu}+1}$ directly without appealing to the finite-size scaling of $\sigma_{\mu \nu}$.

\subsubsection{Odd systems with large interfacial fluctuations}
\label{sec:432}
A second example in which the free-energy exhibits long ranged behaviour occurs in odd systems in the temperature window $T_C>T>T_W$ (often referred to as the soft mode phase \cite{parry1,parry2,boulter,parry3,binder2}). For this case it is most convenient to use (\ref{k+-}) and  (\ref{k+})
\begin{equation}
4 \frac{d^2 \gamma}{dL^2} = \frac{m'_0 \left( \frac{L}{2} \right)^2}{G_0 \left( \frac{L}{2},\frac{L}{2} \right)}
\end{equation}
and recognise that the right hand side is proportional to $\xi_\parallel^{-2}$, where $\xi_\parallel$ is the transverse correlation length characterising fluctuations in the position of the up spin-down spin interface, which wanders `freely' between the two walls. If we recall this correlation length shows finite-size scaling such that \cite{parry2}
\begin{equation}
L \sim \xi_\parallel^\zeta
\end{equation}
with $\zeta$ the roughness exponent we immediately find
\begin{equation}
\gamma(L) \sim L^{2(1-\frac{1}{\zeta})}
\end{equation}
consistent with the fluctuation theory for confined interfaces \cite{vol14}. For purely thermal fluctuations for which $\zeta=\frac{3-d}{2}$ for $d<3$ this reduces to
\begin{equation}
\gamma(L) \sim L^{\frac{-2(d-1)}{3-d}} \mbox{\ \ \ \ \ for $d<3$}
\end{equation}
familiar from the theory of wetting \cite{lipowsky}. The same expression also follows using arguments similar to those given for the case of $T=T_C$ and serve to further illustrate the utility of the results (\ref{K}-\ref{sig}).

\subsubsection{Near the finite-size critical point}
\label{sec:433}
Finally we make some remarks on a third possibility which requires further research. For both odd and even geometries it is natural to expect singularities to emerge as we approach the finite-size critical point occurring at $T_C(L)$ and $h=h_C(L)$. Recall that the locations of these critical points is very different for odd and even symmetries (see Fig.\ 2 and Fig.\ 3). Nevertheless the phase transition occurring at $T_C(L)$ and $h=h_C(L)$ for both these systems is conjectured to belong to the same $d-1$ dimensional bulk Ising universality class \cite{parry1,barber}. However an inspection of the well developed mean-field theories shows that the all important free-energy derivative $\frac{d^2 \gamma}{dL^2}$ has different singularities for odd and even symmetry. In particular $\frac{d^2 \gamma}{dL^2}$ diverges at the finite-size critical point for an even system \cite{evans3} but vanishes (as $T_C \rightarrow T_C(L)^+$) for an odd symmetry \cite{parry1}. This would seem to imply a subtle difference between the behaviour of the correlation functions near the respective critical points (by virtue of (\ref{k+-}) and (\ref{sig})) --- an observation that is perhaps surprising given, as mentioned above, that the universality class of the phase transition is anticipated to be the same for each geometry.

\section{Alternative approaches and some generalizations}
\label{sec:5}
To end our discussion of planar inhomogeneous fluids we present an alternative derivation of the correlation function relations (\ref{PRODUCT},\ref{SUM}) for Landau-type models of the form
\begin{eqnarray}
\Omega[m({\bf r})] & =& \int d{\bf y} \int_{0}^{L} dz \left\{ \frac{1}{2} (\nabla m)^2 + \phi(m) + \delta(z) \phi_1(m) \right. \nonumber \\
& &  + \delta(z-L) \phi_2(m) \Bigr \} \label{HLGW}
\end{eqnarray}
by partial solution of the Ornstein-Zernike equation (\ref{OZ2}) --- a complete solution not being required. Here we emphasis the properties of a kernel or propagator-like function which also emerges by considering the continuum limit of the algebraic relation (\ref{PRODUCT}) (see Sec.\ \ref{sec:52}). Finally we demonstrate that the same algebraic relations amongst the (suitably redefined) $S_{\mu \nu}$ are also obeyed for non-planar inhomogeneous fluids exhibiting a cylindrical or spherical symmetry.

\subsection{The Kernel Function for the Ornstein-Zernike equation}
\label{sec:51}
For the Landau model (\ref{HLGW}) the Ornstein-Zernike equation reduces to \cite{evans1,evans2}
\begin{equation}
\left[ \hat{\cal L}(z_2) + q^2 \right]G(z_1,z_2;{\bf q}) = \delta(z_2-z_1) \label{OZi}
\end{equation}
where the second order linear operator $\hat{\cal L}$ is 
\begin{equation}
\hat{\cal L}(z) = -\frac{\partial^2}{\partial z^2}  + \phi''(m_0(z))  
\end{equation}
The same operator appears in the differential equation for the profile $m_0(z)$ which may be written
\begin{equation}
\hat{\cal L}(z) m_0'(z) =0 \label{ELii}
\end{equation}
and will be required later. To proceed we first define the function
\begin{equation}
{\cal K}(z_1,z_2;{\bf q}) = \frac{\partial}{\partial z_2} \log G(z_1,z_2;{\bf q}) \label{riccati}
\end{equation}
which from (\ref{OZi}) satisfies the non-linear equation (for $z_1 \ne z_2$)
\begin{equation}
\frac{d {\cal K}}{dz_2} = \phi''(m(z_2))+q^2 -{\cal K}^2  \label{k}
\end{equation}
Assuming (\ref{k}) has a family of solutions $k(z_2,{\bf q};a)$, parameterized by $a$, then
\begin{equation}
{\cal K}(z_1,z_2;{\bf q}) = \left\{ \begin{array}{ll} k(z_2,{\bf q};a_1) & \mbox{\ \ \ \ \ for $z_2>z_1$} \\ k(z_2,{\bf q};a_2) & \mbox{\ \ \ \ \ for $z_2<z_1$} \end{array} \right.
\end{equation}
and the $\delta$ function in (\ref{OZi}) implies the boundary condition
\begin{equation}
k(z_1,{\bf q};a_1) - k(z_1,{\bf q};a_2) = -\frac{1}{G(z_1,z_1;{\bf q})}
\end{equation}
Without loss of generality this can be considered an equation for $a_2$ and implies that $a_2=a_2(z_1,{\bf q})$. Consequently, 
\begin{equation}
{\cal K}(z_1,z_2;{\bf q}) = {\cal K}(z_2;{\bf q}) \mbox{\ \ \ \ \ for $z_1 \le z_2$} \label{Key}
\end{equation}
As we shall see the existence of ${\cal K}$ is central to the derivation of the correlation function relations. Integrating (\ref{riccati}) and remembering (\ref{Key}) we find (for $z_1<z_2$)
\begin{equation}
G(z_1,L;{\bf q})-G(z_1,z_2;{\bf q}) = \int_{z_2}^L dz' {\cal K}(z';{\bf q}) G(z',z_1;{\bf q}) 
\end{equation}
This becomes in real space (via the convolution theorem)
\begin{equation}
G({\bf r}_1,{\bf r}_2) = G({\bf r}_1,({\bf y}_2,L))+ \int d{\bf r}' G({\bf r}_1,{\bf r}') K^-({\bf r}',{\bf r}_2) \label{one2}
\end{equation}
where $z_1<z_2$ and $K^-$ can be thought of as an advanced propagator
\begin{equation}
K^-({\bf r}',{\bf r}_2) \equiv - \theta(z'-z_2) K(z';{\bf y}_2-{\bf y}')
\end{equation}
Here $K(z;{\bf y})$ is the Fourier transform of the kernel ${\cal K}(z;{\bf q})$. In this way it is possible to interprete the correlation function at positions ${\bf r}_1$ and ${\bf r}_2$ (say) in a standard way i.\ e.\ making analogy between position  $z$ and time, $G({\bf r}_1,{\bf r}')$ at early `time' $z=z_1$ can be propogated to later time $z=z_2$ making use of information from the `future' ($z'>z_2$) only.

The `product' rule follows directly from the integral relation (found from (\ref{riccati}) and (\ref{Key}))
\begin{equation}
G(z_1,b;{\bf q}) = G(z_1,a;{\bf q}) \exp \left[ \int_a^b dz {\cal K}(z;{\bf q}) \right] \label{three}
\end{equation}
valid for all $b \ge a \ge z_1$. Setting $z_1=a$ and choosing $c$ satisfying $a \le c \le b$ we can write
\begin{eqnarray}
G(a,b;{\bf q}) & = & G(a,a;{\bf q}) \exp \left[ \int_a^b dz {\cal K}(z;{\bf q}) \right] \nonumber \\
& = & G(a,a;{\bf q}) \exp \left[ \int_a^c dz {\cal K}(z;{\bf q}) \right] \nonumber \\ 
& & \times \exp \left[ \int_c^b dz {\cal K}(z;{\bf q}) \right] \nonumber \\
& = & G(a,a;{\bf q}) \frac{ G(a,c;{\bf q})}{ G(a,a;{\bf q})} \frac{ G(b,c;{\bf q})}{ G(c,c;{\bf q})} \label{op}
\end{eqnarray}
where we have used (\ref{three}) extensively. The final equation is nothing more than (\ref{PRODUCT}).

The second correlation function relation can be found from the explicit solution of (\ref{k}) when $q=0$
\begin{equation}
{\cal K}_0 = \frac{d}{dz} \left[ \log Y(z) \right] \label{q}
\end{equation}
where $\frac{d^2 Y}{dz^2}= \phi''(m_0(z)) Y(z)$. From the definition (\ref{riccati}) one finds (for $z_2,z_3 \ge z_1$)
\begin{eqnarray}
\frac{G_0(z_1,z_2)}{G_0(z_1,z_3)} & = & \frac{Y(z_2)}{Y(z_3)}\nonumber \\ 
\Rightarrow G_0(z_1,z_2) & = & Y(z_2) X(z_1)
\end{eqnarray}
for some function $X(z)$. Enforcing the Ornstein-Zernike conditions
\begin{equation}
\hat{\cal L}(z_1) G_0(z_1,z_2) = \hat{\cal L}(z_2) G_0(z_1,z_2) = \delta(z_1-z_2)
\end{equation}
we see that the functions $X$ and $Y$ are related by
\begin{equation}
X(z) = \alpha Y(z) +\beta m_0'(z) 
\end{equation}
for constant $\alpha$ and $\beta$, due to the linear nature of the operator $\hat{\cal L}(z)$ and (\ref{ELii}). This is precisely equivalent to the linear relation (\ref{kernellink}) which we deduced as a necessary consequence of the second correlation function identity (\ref{SUM}).

\subsection{The continuum limit}
\label{sec:52}
The propagator approach also emerges from the more general stiffness-matrix formalism by taking the continuum limit of (\ref{PRODUCT}). Consider two boundary planes located at $z=a$ and $z=b$ (with $a<b$) and divide the region between them by $n$ other planes separated by a constant distance $h$. Combining `product' rules (\ref{PRODUCT}) defined for each consecutive set of two planes and the boundary at $z=b$ one finds
\begin{equation}
G(a,b;{\bf q}) = \frac{G(a,z_1;{\bf q}) G(z_1,z_2;{\bf q}) \ldots G(z_n,b;{\bf q})}{G(z_1,z_1;{\bf q}) G(z_2,z_2;{\bf q}) \ldots G(z_n,z_n;{\bf q})} \label{ratios}
\end{equation}
Setting $z_0=a$ and $z_{n+1}=b$, (\ref{ratios}) can be rewritten as
\begin{eqnarray}
\frac{G(a,b;{\bf q})}{G(a,a;{\bf q})} & = & \frac{\prod_{i=0}^{i=n} G(z_i,z_i+h;{\bf q})}{\prod_{i=0}^{i=n} G(z_i,z_i;{\bf q})} \nonumber \\
& = & \exp [ \Sigma_{i=0}^{n} \{ \log G(z_i,z_i+h;{\bf q}) \nonumber \\
& &  - \log G(z_i,z_i;{\bf q}) \} ] \nonumber \\
& = & \exp \left[ \Sigma_{i=0}^{n} h \left\{  \frac{ \log G(z_i,z_i+h;{\bf q})}{h} \right. \right. \nonumber \\
& & \left. \left.  - \frac{\log G(z_i,z_i;{\bf q})}{h} \right\} \right] 
\end{eqnarray}
Taking the continuum limit $h \rightarrow 0$, $n \rightarrow \infty$ we find that
\begin{equation}
G(a,b;{\bf q}) = U(a,b;{\bf q}) G(a,a;{\bf q}) \mbox{\ \ \ \ \ for $a<b$} \label{result1}
\end{equation}
where
\begin{equation}
U(a,b;{\bf q}) \equiv \exp \left[ \int_{a}^{b}dz \, {\cal K}(z;{\bf q}) \right] 
\end{equation}
and ${\cal K}(z;{\bf q})$ is some unknown function. For an infinitesimal displacement $dz_2$,
\begin{eqnarray}
U(z_1,z_2+dz_2;{\bf q}) & = & \exp \left[ \int_{z_1}^{z_2+dz_2} dz \, {\cal K}(z;{\bf q}) \right] \nonumber \\
& \simeq & \exp [ {\cal K}(z_2;{\bf q}) dz_2] U(z_1,z_2;{\bf q}) \nonumber \\
& \simeq & \left[ 1 + {\cal K}(z_2;{\bf q}) dz_2 \right]  U(z_1,z_2;{\bf q})
\end{eqnarray}
which implies
\begin{eqnarray}
\frac{U(z_1,z_2+dz_2)-U(z_1,z_2)}{dz_2} & = & {\cal K}(z_2;{\bf q}) U(z_1,z_2;{\bf q}) \nonumber \\
\Rightarrow \frac{\partial}{\partial z_2} U(z_1,z_2;{\bf q}) & = & {\cal K}(z_2;{\bf q}) U(z_1,z_2;{\bf q}) \label{uevolve}
\end{eqnarray}
taking the limit $dz_2\rightarrow 0$. It is intriguing that (\ref{uevolve}) is very similar to the Schr\"{o}dinger equation for the time evolution operator, with the Hamiltonian time dependent but self commuting at different times \cite{sakurai}. 

Multiplying (\ref{uevolve}) by $G(z_1,z_1;{\bf q})$ and then using (\ref{result1})
\begin{eqnarray}
\frac{\partial}{\partial z_2} U(z_1,z_2;{\bf q}) G(z_1,z_1;{\bf q}) & = & {\cal K}(z_2;{\bf q}) U(z_1,z_2;{\bf q}) \nonumber \\
& & \times G(z_1,z_1;{\bf q}) \nonumber  \\
\Rightarrow \frac{\partial}{\partial z_2} G(z_1,z_2;{\bf q}) & = & {\cal K}(z_2;{\bf q}) G(z_1,z_2;{\bf q}) 
\end{eqnarray}
which is equivalent to (\ref{riccati}) with (\ref{Key}), and ${\cal K}(z;{\bf q})$ is, of course, just the kernel function.

Thus we have shown that (\ref{PRODUCT}) implies via its continuum limit the existence of a kernel function ${\cal K}(z;{\bf q})$. In turn the kernel function can be understood as being an advanced propagator for correlation functions in real space. 

\subsection{Non-planar geometries}
\label{sec:53}
The properties of inhomogeneous fluids confined in {\it non-planar} geometries has also attracted attention in recent years. Perhaps the most important of these is the ease of fluid adsorption in cylindrical systems \cite{evans4} (as idealised models of porous materials). Despite continued interest we are not aware of any discussion of correlation function structure in such systems and to complete our article we make some remarks which follows from those mentioned in Sec.\ \ref{sec:51}.

Consider then a fluid confined in an infinitely long cylinder ($-\infty<z<\infty$) of radius $R$ (and connected to an external reservoir of particles). The grand-potential density functional is taken to be the Landau-type model
\begin{eqnarray}
\Omega[m] & = & \int_{-\infty}^{\infty}dz \int_0^{2\pi} d\phi \int_0^R rdr \left\{ \frac{1}{2} (\nabla m)^2 +\phi(m) \right.  \nonumber \\
& & +\delta(r-R)\phi_1(m) \Bigl \}
\end{eqnarray}
analogous to (\ref{HLGW}). The connected correlation function $G({\bf r}_1,{\bf r}_2)$ only depends on the relative angle $\phi_2-\phi_1\equiv \phi_{21}$ and azimuthal distance $z_2-z_1 \equiv z_{21}$ between the particles as well as the radial distances $r_1$ and $r_2$. To exploit this we first define a `zeroth' moment analogous to (\ref{goo})
\begin{equation}
G_0(r_1,r_2)=\int_{-\infty}^{\infty}dz_{21} \int_0^{2\pi}d\phi_{21}G({\bf r}_1,{\bf r}_2)
\end{equation}
The Ornstein-Zernike equation is
\begin{equation}
\left[ \nabla_2^2+\phi''(m(r_2;R)) \right] G({\bf r}_2,{\bf r}_1)=\delta({\bf r}_2-{\bf r}_1)
\end{equation}
where $m(r;R)$ is the equilibrium profile satisfying the Euler-Lagrange equation
\begin{equation}
\frac{1}{r} \frac{\partial}{\partial r} \left(r \frac{\partial}{\partial r} \right)m(r;R) = \phi'(m(r;R))
\end{equation}
together with a boundary condition at $r=R$. The differential equation for $G({\bf r}_1,{\bf r}_2)$ can be integrated with respect to $z_{21}$ and $\phi_{21}$ to yield (for $r_1 \ne r_2$)
\begin{equation}
\left[ -\frac{1}{r_2} \frac{\partial}{\partial r_2} \left(r_2 \frac{\partial}{\partial r_2} \right)+ \phi''(m(r_2;R)) \right] G_0(r_1,r_2) = 0 
\end{equation}
which we need to solve. The point to notice here is that the partial derivative $\frac{\partial m}{\partial R}(r;R)$ satisfies the same linear differential equation
\begin{equation}
\left[ -\frac{1}{r_2} \frac{\partial}{\partial r_2} \left(r_2 \frac{\partial}{\partial r_2} \right)+ \phi''(m(r_2;R)) \right] \frac{\partial m}{\partial R}(r_2;R) = 0 
\end{equation}
As a consequence it is natural to define the `structure' factor matrix elements as
\begin{equation}
S_{\mu \nu}(0) \equiv \frac{G_0(r_\mu,r_\nu)}{\frac{\partial m}{\partial R}(r_\mu;R) \frac{\partial m}{\partial R}(r_\nu;R)}
\end{equation}
which, we anticipate, satisfy the same algebraic conditions (\ref{PRODUCT}) and (\ref{SUM}) as the planar system. This is borne out by the explicit solution
\begin{eqnarray}
G_0(r_1,r_2) & = & \frac{\partial m}{\partial R}(r_1;R) \frac{\partial m}{\partial R}(r_2;R) \Bigl ( \alpha \nonumber \\
& & + \int_{\max (r_1,r_2)}^R dr r^{-1} \left( \frac{\partial m}{\partial R} \right)^{-2} \Bigr )
\end{eqnarray}
where $\alpha$ is determined by a simple boundary condition. In terms of the structure factors we note that
\begin{equation}
S_{\mu \nu}(0) = S_{\mu \mu}(0) \mbox{\ \ \ \ \ $\forall r_\nu \ge r_\mu$} 
\end{equation}
which is analogous to the semi-infinite solution (\ref{semi}) for planar systems. Indeed for fluid adsorption on the outside of a cylinder the Landau theory result is 
\begin{eqnarray}
G_0(r_1,r_2) & = & \frac{\partial m}{\partial R}(r_1;R) \frac{\partial m}{\partial R}(r_2;R) \Bigl( \alpha'  \nonumber \\
& & + \int_R^{\min (r_1,r_2)} dr r^{-1} \left( \frac{\partial m}{\partial R} \right)^{-2} \Bigr)
\end{eqnarray}
which is precisely of the form (\ref{LPl}).

It is a simple matter to repeat the analysis for fluid adsorption in spherically symmetric systems (coordinates $r,\theta,\phi$) where the appropriate zeroth moment is
\begin{equation}
G_0(r_1,r_2)=\int_{0}^{\pi} \sin(\theta_{12}) d\theta_{12} \int_0^{2 \pi}d\phi_{21}G({\bf r}_1,{\bf r}_2)
\end{equation}
where $\theta_2-\theta_1 \equiv \theta_{12}$. In fact, if we restrict our attention to adsorption at a single wall (with $z \leftrightarrow r$ and $z_0 \leftrightarrow R$), or outside cylinders and spheres the general solution can be written
\begin{eqnarray}
G_0(r_1,r_2) &=& \frac{\partial m}{\partial R}(r_1;R) \frac{\partial m}{\partial R}(r_2;R) \Bigl( {\rm constant} \nonumber \\
& & + \int_R^{\min (r_1,r_2)} dr r^{-d'} \left( \frac{\partial m}{\partial R} \right)^{-2} \Bigr)
\end{eqnarray}
where $d'=0,1,2$ for walls, cylinders and spheres respectively. We can interpret $d'$ as the dimension of the boundary surface (two for three dimensional systems) minus the number of dimensions in which the systems is unbounded. For all these systems the structure factors satisfy (\ref{semi}). For the more general case of thin film geometries and adsorption between two concentric cylinders or spheres the structure factors can be shown \cite{us} to satisfy the general algebraic conditions (\ref{PRODUCT}) and (\ref{SUM}).

\section{Conclusions}
\label{sec:6}
To complete our article we summarise our main results and make some remarks about possible future work.

\newcounter{c}
\begin{list}{(\roman{c})}{\usecounter{c}}
\item The stiffness-matrix formalism for calculating correlation functions $G(z_\mu,z_\nu;q)$ in planar inhomogeneous fluids modelled by local variational models naturally leads to two identities relating this function at three arbitrary positions $z_1 \le z_2 \le z_3$. Taken together they restrict the form of the zeroth moment $G_0(z_\mu,z_\nu)$ and allow us to define dimensionless scaled variables
\begin{equation}
\sigma_{\mu \nu} \equiv  \frac{4}{k_BT} \frac{d^2\gamma}{dL^2} \frac{G_0(z_\mu,z_\nu)}{m_0'(z_\mu)m_0'(z_\nu)} 
\end{equation}
for thin film geometries exhibiting even or odd reflection symmetries. The cross term $\sigma_{12}$ is solved for explicitly as a universal function of $\sigma_{11}$ and $\sigma_{22}$.
\item For even systems and particular choices of $z_1,z_2$ and $z_3$ our results are consistent with previously derived exact expressions for fluid adsorption at purely repulsive hard walls \cite{henderson} with arbitrary fluid-fluid interactions.
\item Using our results we have been able to rederive non-trivial scaling expressions for the long-ranged (power law) finite-size contribution to the excess free-energy $\gamma(L)$

(a) at the bulk critical point

(b) in the soft mode phase for a geometry with competing surface fields (odd systems)

There is a hint in our analysis that the correlation function may behave somewhat differently near the capillary critical point of odd and even symmetric systems.
\item While the correlation function identities may also be derived directly, for Landau-type models, by explicit solution of the Ornstein-Zernike equation, it is unlikely that the general character of the relations would have been spotted using this method. This serves to illustrate the utility of the stiffness-matrix formalism which has been previously used to derive the stiffness-matrix free-energy relation in the theory of wetting \cite{parry4}. One may add here that while most of our analysis has been restricted to $q=0$ one may also discuss the structure of higher moments of the correlation function. For example, consider the position dependent transverse correlation length $\xi_\parallel(z_\mu,z_\nu)$, defined via the asymptotic expansion (for $q \rightarrow 0$)
\begin{eqnarray}
G(z_\mu,z_\nu;q) & = & G_0(z_\mu,z_\nu) [1-\xi_\parallel^2(z_\mu,z_\nu) q^2 \nonumber \\
& &  + O(q^4)] 
\end{eqnarray}
Then, from the `product' identity (\ref{PRODUCT}) we immediately derive the elegant correlation length relation
\begin{equation}
\xi_\parallel^2(z_1,z_2) + \xi_\parallel^2(z_2,z_3)= \xi_\parallel^2(z_2,z_2)+ \xi_\parallel^2(z_1,z_3)
\end{equation}
for $z_1 \le z_2 \le z_3$. Using this identity and other stiffness-matrix relations it is possible to show \cite{us} that the three correlation lengths $\xi_\parallel^2(\frac{L}{2},\frac{L}{2}), \xi_\parallel^2(0,\frac{L}{2})$ and $\xi_\parallel^2(0,L)$ all diverge in precisely the same manner as $L \rightarrow \infty$ at $T=T_C$ (even and odd systems) and $T_C>T>T_W$ (odd systems) in zero bulk field.
\item Finally, we believe that it would be worthwhile developing the stiffness-matrix formalism for other variational functionals, for example, one could consider local models of two or more density variables such as arise in systems with $N$-component vector order parameters. Within the same framework it would be possible to study local entropy-type functionals $\Omega[m({\bf r}),\epsilon({\bf r})]$ of the magnetization and the energy density which have been forwarded recently as better candidates for modelling critical behaviour \cite{mikheev}. A preliminary analysis of these models reveals analogues of the `product' rule (\ref{PRODUCT}) although further work is required to find whether the identity (\ref{SUM}) also generalizes.
\end{list}

\acknowledgements
A.\ O.\ Parry is very grateful for Prof.\ E.\ Domay for his hospitality at the Weizmann Institute, Israel, where this work was begun and we acknowledge financial support from the Engineering and Physical Sciences Research Council, United Kingdom.

\appendix
\section{A separable binding potential}
Let us denote $m_\pi^{(N)}(z; \{ \ell_\alpha \})$ as the planar profile which minimizes $\Omega[m]$ subject to the crossing constraints (\ref{crossing}). The binding potential is determined by the relation (dropping the implicit $\{z_\alpha \}$ dependence)
\begin{eqnarray*} 
& W_N( \{ \ell_\alpha \}) = \phi_1( m_\pi^{(N)}(0; \{ \ell_\alpha \}) ) + \phi_2( m_\pi^{(N)}(L; \{ \ell_\alpha \}) ) & \\
& + \int_0^L dz {\cal L}^{(b)} \left( m_\pi^{(N)} (z; \{ \ell_\alpha \}), \partial_z m_\pi^{(N)}(z; \{ \ell_\alpha \}) \right) &
\end{eqnarray*}
neglecting constant terms independent of the collective coordinates. This is profitably rewritten
\begin{eqnarray}
& W_N( \{ \ell_\alpha \})= \phi_1( m_\pi^{(N)}(0; \{ \ell_\alpha \}) ) \nonumber \\
& + \phi_2( m_\pi^{(N)}(L; \{ \ell_\alpha \}) ) + \sum_{n=0}^N \int_{\ell_n}^{\ell_{n+1}} dz & \nonumber \\
& {\cal L}^{(b)} \left( m_\pi^{(N)} (z; \{ \ell_\alpha \}), \partial_z m_\pi^{(N)}(z; \{ \ell_\alpha \}) \right) & \label{a2} 
\end{eqnarray}
where we have defined $\ell_0=0$ and $\ell_{N+1}=L$. The profile itself satisfies the second order Euler-Lagrange equation
\begin{equation}
\frac{\partial}{\partial z} \left( \frac{\partial {\cal L}^{(b)}}{\partial \left( \partial_z m_\pi^{(N)} \right)} \right) = \frac{\partial{\cal L}^{(b)}}{\partial m_\pi^{(N)}}
\end{equation}
so that for $\ell_\mu <z<\ell_{\mu+1}$ the solution is completely determined by the boundary conditions $ m_\pi^{(N)} \left(\ell_\mu; \{ \ell_\alpha \} \right)=m^X_\mu$ and $m_\pi^{(N)} \left(\ell_{\mu+1}; \{ \ell_\alpha \} \right)=m^X_{\mu+1}$. Consequently we can write (for $\ell_\mu \le z \le \ell_{\mu+1}$)
\begin{equation}
m_\pi^{(N)} \left(z; \{ \ell_\alpha \} \right)=F_\mu(z-\ell_\mu,\ell_{\mu+1}-\ell_\mu) 
\end{equation}
for all $\mu$ and with $F_\mu$ some appropriate function which we need not determine. Substituting into (\ref{a2}) then yields
\begin{equation}
W_N(\{ \ell_\mu \}) = \sum_{n=1}^{N+1} V_n(\ell_n-\ell_{n-1})
\end{equation}
exhibiting the required separable properties.

\section{Connection with the free-energy}
Consider odd systems. We start by using (\ref{hendy}) so that (\ref{K}) can be rewritten as
\begin{equation}
K_+ = \frac{4}{c^2} G_0(0,L)
\end{equation}
and using (\ref{hendy}) again (with $z=\frac{L}{2}$)
\begin{equation}
K_+ = \frac{G_0(0,L)}{G_0(0,\frac{L}{2})^2} m'_0(\frac{L}{2})^2
\end{equation}
However from the `product' relation (\ref{PRODUCT})
\begin{equation}
G_0(0,\frac{L}{2})^2 =G_0(0,L) G_0(\frac{L}{2},\frac{L}{2})
\end{equation}
so that
\begin{equation}
K_+ = \frac{m'_0(\frac{L}{2})^2}{G_0(\frac{L}{2},\frac{L}{2})} \label{ak+}
\end{equation}
which is (\ref{k+}). The right hand side of (\ref{ak+}) is the inverse of the structure factor matrix element for a surface of fixed magnetization $m^X=0$ (located at the centre of the thin film on the average). If we calculate the binding potential $W_1(\ell)$ for a {\it single} surface of fixed magnetization $m^X=0$ we can identify
\begin{equation}
\gamma(L) = W_1(\frac{L}{2})
\end{equation}
However from the stiffness-matrix formalism with $N=1$ we also find
\begin{eqnarray}
\frac{m'_0(\frac{L}{2})^2}{G_0(\frac{L}{2},\frac{L}{2})} & = & W_1''(\ell) \mbox{ \ \ \ \ \ for $\ell=\frac{L}{2}$} \nonumber \\
& = & 4 \frac{d^2 W_1 \left( \frac{L}{2} \right)}{dL^2} 
\end{eqnarray}
valid in the soft mode phase $T \ge T_W$ where $z=\frac{L}{2}$. This completes the proof.

A similar approach can be used for even systems and therefore we find generally
\begin{equation}
K_\pm = 4 \frac{d^2 \gamma(L)}{dL^2}
\end{equation}

\end{document}